\RequirePackage{fix-cm}
\documentclass[smallextended]{svjour3}       
\smartqed  
\usepackage{graphicx}
\usepackage{latexsym}
\usepackage{amssymb}
\usepackage{amsmath}
\usepackage{amsmath,empheq}
\usepackage{booktabs}
\usepackage[export]{adjustbox}
\usepackage{subfig}
\usepackage{enumerate}
\usepackage{multicol}
\usepackage{rotating}
\usepackage{pdflscape}
\DeclareMathSizes{12}{8}{7}{4}
\usepackage{nicefrac}
\usepackage{mathtools}
\usepackage{svg}
\usepackage{sidecap}
\usepackage{siunitx}
\usepackage{placeins}
\usepackage{comment}
\usepackage{sidecap}

\usepackage{xargs}                      

\usepackage[sc]{mathpazo}
\usepackage{algorithm}
\usepackage[noend]{algpseudocode}
\makeatletter
\renewcommand\fs@ruled{%
  \def\@fs@cfont{\itshape}%
  \let\@fs@capt\floatc@plain%
  \def\@fs@pre{\hrule height.8pt depth0pt \kern2pt}
  \def\@fs@post{}
  \def\@fs@mid{\kern2pt\hrule\kern2pt}
  \let\@fs@iftopcapt\iffalse}
\makeatother
\makeatletter
\renewcommand{\ALG@name}{\small\textbf{Algorithm}}
\makeatother
\usepackage{etoolbox}

\usepackage{mwe,tikz}\usepackage[percent]{overpic}

\usepackage[sort,compress]{cite}
\usepackage{multirow}
\usepackage{makecell}
\usepackage{textcomp}
\usepackage{paralist}
\usepackage{cancel}
\usepackage{color, colortbl}
\usepackage{longtable}
\usepackage{graphbox}
\usepackage{cases} 

\usepackage[labelfont=bf,textfont=it,belowskip=0pt,aboveskip=5pt,tableposition=top]{caption}
\usepackage[colorlinks=true,
citecolor=blue,
filecolor=black,
linkcolor=blue,
urlcolor=blue]{hyperref}
\usepackage{lineno}

\graphicspath{{figures/}{tables/}}

\newcounter{runidnum}
\newcommand{\runid}{\stepcounter{runidnum}\#\therunidnum}

\newcommand{\figref}[1]{Fig.~\ref{#1}}
\newcommand{\tabref}[1]{Tab.~\ref{#1}}
\newcommand{\secref}[1]{\S\ref{#1}}

\newcommand{\nf}[2] {\nicefrac{#1}{#2}}

\newcolumntype{R}{>{\columncolor{gray!20}}r}
\newcolumntype{C}{>{\columncolor{gray!20}}c}
\newcolumntype{L}{>{\columncolor{gray!20}}l}
\newcolumntype{G}{>{\columncolor{colorA!30}}l}

\newcommand{\mpPatientBrain}[1][]{\ifthenelse { \equal {#1} {} }
    { \vect{m}_D }   
    { \vect{m}_{D,#1} } }  

\newcommand{\mpWarpedAtlasBrain}[1][]{\ifthenelse { \equal {#1} {} }
    { \vect{m}_A^{(1,1)} }   
    { \vect{m}_{A,#1}^{(1,1)} }}   

\newcommand{\D}[1]{\ensuremath{\mathcal{#1}}}                           

\newcommand{\idiv}{\ensuremath{\nabla\cdot}}
\newcommand{\igrad}{\ensuremath{\nabla}}

\renewcommand{\d}[1]{\mathop{}\!\mathrm{d}#1}

\newcommand{\dx}{\d{\vect{x}}}

\newcommand{\p} {\partial}

\newcommand{\vect}[1]{\boldsymbol{#1}} 
\newcommand{\mat}[1]{\boldsymbol{#1}}  

\newcommand{\bipa}{\begin{inparaenum}[(\itshape i\upshape)]}
\newcommand{\eipa}{\end{inparaenum}}

\newcommand{\bipasub}{\begin{inparaenum}[(\itshape a\upshape)]}
\newcommand{\eipasub}{\end{inparaenum}}

\newcommand{\ipoint}[1]{\textit{\textbf{\color{black}#1}}}

\newcommand{\defeq}{\ensuremath{\mathrel{\mathop:}=}}

\definecolor{sred}{cmyk}{0.01,0.98,0,0.2} 
\reversemarginpar
\newcommand{\mmargin}[1]{{\marginpar{\em\tiny #1}}}\renewcommand{\mmargin}[1]{}

\definecolor{C0}{rgb}{0.000,0.447,0.741}
\definecolor{C1}{rgb}{0.850,0.325,0.098}
\definecolor{C2}{rgb}{0.929,0.694,0.125}
\definecolor{C3}{rgb}{0.494,0.184,0.556}
\definecolor{C4}{rgb}{0.466,0.674,0.188}
\definecolor{C5}{rgb}{0.301,0.745,0.933}
\definecolor{C6}{rgb}{0.635,0.078,0.184}
\definecolor{C7}{rgb}{0.887,0.465,0.758}
\definecolor{C8}{rgb}{0.496,0.496,0.496}

\definecolor{anthrazit}{RGB}{62,68,76}
\definecolor{mittelblau}{RGB}{0,81,158}
\definecolor{helllblau}{RGB}{0,190,255}

\definecolor{mdtRed}{RGB}{180,0,0}

\algrenewcommand{\alglinenumber}[1]{\footnotesize\color{anthrazit}{{#1}}}
\algrenewcommand{\algorithmicrequire}{\textbf{Input:}}
\algrenewcommand{\algorithmicensure}{\textbf{Output:}}

\algrenewcommand{\textproc}{}

\algnewcommand{\Break}{\textbf{break}}
\algnewcommand{\Continue}{\textbf{continue}}
\algnewcommand{\True}{\textbf{true}}
\algnewcommand{\False}{\textbf{false}}
\algnewcommand{\Null}{\textbf{null}}
\algrenewcommand{\algorithmicend}{\textbf{end}}
\algrenewcommand{\algorithmicdo}{\textbf{do}}
\algrenewcommand{\algorithmicwhile}{\textbf{while}}
\algrenewcommand{\algorithmicfor}{\textbf{for}}
\algrenewcommand{\algorithmicforall}{\textbf{for all}}
\algrenewcommand{\algorithmicloop}{\textbf{loop}}
\algrenewcommand{\algorithmicrepeat}{\textbf{repeat}}
\algrenewcommand{\algorithmicuntil}{\textbf{until}}
\algrenewcommand{\algorithmicprocedure}{\textbf{procedure}}
\algrenewcommand{\algorithmicfunction}{\textbf{func}}
\algrenewcommand{\algorithmicif}{\textbf{if}}
\algrenewcommand{\algorithmicthen}{\textbf{then}}
\algrenewcommand{\algorithmicelse}{\textbf{else}}
\algrenewcommand{\algorithmicreturn}{\textbf{return}}
\algnewcommand{\algorithmicforever}{\textbf{for ever}}
\algdef{S}[FOR]{ForEver}{\algorithmicforever\ \algorithmicdo}
\algnewcommand{\algorithmicgoto}{\textbf{go to}}
\algnewcommand{\Goto}[1]{\algorithmicgoto\ \cref*{#1}}
\algnewcommand{\funccall}[1]{\textbf{\color{darkgray}#1}}

\makeatletter
\g@addto@macro\ALG@beginalgorithmic{\small}
\makeatother

\algrenewcomment[2][.3\linewidth]{\hfill\leavevmode\hfill\makebox[#1][l]{$\#$ \emph{\normalfont\color{darkgray}#2}}}

\definecolor{pastGcolrA}{RGB}{4,92,155}
\definecolor{pastGcolrB}{RGB}{6,140,117}
\definecolor{pastGcolrC}{RGB}{122,211,227}
\definecolor{pastGcolrD}{RGB}{12,163,171}
\definecolor{pastGcolrE}{RGB}{124,180,212}

\definecolor{colorA}{RGB}{189,201,225}
\definecolor{babyblue}{RGB}{29,162,220}
\definecolor{anthrazit}{RGB}{0,153,229}
\definecolor{red}{RGB}{240,85,97}
\definecolor{colorB}{RGB}{103,169,207}
\definecolor{colorC}{RGB}{ 28,144,153}
\definecolor{colorD}{RGB}{  1,108, 89}
\definecolor{anthrazit}{RGB}{62,68,76}
\definecolor{mittelblau}{RGB}{0,81,158}
\definecolor{hellblau}{RGB}{0,190,255}

\definecolor{c1color1}{RGB}{125,97,234}
\definecolor{c1color2}{RGB}{100,200,214}
\definecolor{c1color3}{RGB}{209,241,154}
\definecolor{c1color4}{RGB}{255,20,20}
\definecolor{c0color1}{RGB}{243,77,250}
\definecolor{c0color2}{RGB}{121,173,252}
\definecolor{dcolor}{RGB}{255,255,228}
\definecolor{segED}{RGB}{64,64,64}
\definecolor{segWM}{RGB}{192,192,192}
\definecolor{segGM}{RGB}{160,160,160}
\definecolor{segCSF}{RGB}{255,255,255}

\definecolor{pastelgreen}{RGB}{60,179,113}

\usepackage{xcolor}
\usepackage{graphicx}
\usepackage{rotating}
\usepackage{lscape}

\begin{document}

\title{Calibration of Biophysical Models for tau-Protein Spreading in Alzheimer's Disease from PET-MRI
}

\titlerunning{Calibration of Biophysical Models for tau Spreading in AD}        

\author{Klaudius Scheufele \and
        Shashank Subramanian \and
        George Biros 
}


\institute{K. Scheufele \at
              Oden Institute, University of Texas at Austin, Texas,\\
              \email{klaudius.scheufele@gmail.com}           
           \and
           S. Subramanian \at
                         Oden Institute, University of Texas at Austin, Texas,\\
                         \email{shashanksubramanian@utmail.utexas.edu}
          \and
          G. Biros \at
                        Oden Institute, University of Texas at Austin, Texas,\\
                        \email{biros@oden.utexas.edu}
}
\date{Received: date / Accepted: date}

\maketitle

\begin{abstract}

Aggregates of misfolded tau proteins (or just  \emph{``tau''} for brevity) play a crucial role in the progression of Alzheimer's disease (AD) as they correlate with cell death and accelerated tissue atrophy. Longitudinal positron emission tomography (PET) scans can be used quantify the extend of abnormal tau spread. Such PET-based image biomarkers are a promising technology for  AD diagnosis and prognosis. Here, we propose to calibrate an organ-scale biophysical mathematical model using longitudinal PET scans to extract characteristic growth patterns and spreading of tau. The biophysical model is a reaction-advection-diffusion partial differential equation (PDE) with only two scalar unknown parameters, one representing the  spreading (the diffusion part of the PDE) and the other one the growth of tau (the reaction part of the PDE). The advection term captures tissue atrophy and is obtained from diffeomorphic registration of longitudinal  magnetic resonance imaging (MRI) scans. We describe the method, present a numerical scheme for the calibration  of the growth and spreading parameters, perform a sensitivity study using synthetic data, and we perform a preliminary evaluation on clinical scans from the ADNI dataset.
We study whether such model calibration is possible and investigate the sensitivity of such calibration to the time between consecutive scans and the presence of atrophy. Our findings show that despite using only two calibration parameters, the model can reconstruct clinical scans quite accurately.
We discovered that small time intervals between scans and the presence of background noise create difficulties. Furthermore, we evaluate the sensitivity of the model calibration to the offset considered for background noise subtraction. Our reconstructed model fits the data well, yet the study on clinical data also reveals shortcomings of the simplistic model. Interestingly, the values of the model parameters have significant variability across patients, an indication that these parameters could be useful biomarkers.

\end{abstract}

\section{Introduction}\label{sec:intro}

Alzheimer's disease is the sixth highest leading cause of death in the US (\href{https://alz.org}{alz.org}).
Its complex evolution is thought to be closely related to the formation and spreading of abnormal proteinaceous assemblies in the nervous system. In particular, the toxic misfolding of $\beta$-amyloid (A$\beta$) and tau proteins are believed to be key factors driving the progression of Alzheimer's disease~\cite{Jucker:2013a,Walker:2013,Jucker:2018a}.
These corruptive protein templates incite a chain reaction of misfolding, by inducing their anomalous structure on benign molecules. Subsequent growth, fragmentation and further spreading of such toxic proteins hampers proper function of the nervous system, leads to accelerated tissue atrophy, necrosis, and ultimately causes death~\cite{Brettschneider:15a,Ossenkoppele:2016,La-Joie:2020,Rowe:2013}.

Tau aggregates are primarily found in the axon bundle and rapidly spread along neuronal pathways to distant locations, but also invade the extracellular space~\cite{Jucker:2011a}. Understanding the distinct spatiotemporal growth patterns and original seeding of corrupted protein templates is imperative in developing new treatment protocols and can reveal important complimentary information to aid diagnosis and overall efficacy of therapy. Longitudinal PET scans---using the F-AV-1451 tracer, also referred to as tauvid\footnote{http://pi.lilly.com/us/tauvid-uspi.pdf}---can image tau spreading and lead to improved diagnosis and prognosis.
Many groups are designing image-analysis algorithms for this purpose~\cite{James:2015,Lowe:2016,Smith:2016,Scholl:2017}. Here we propose a complementary approach:
We employ a PDE model of tau propagation and calibrate its parameters using longitudinal PET scans.
In clinical practice, tau-PET is acquired in intervals of 8 to 15 months with varying, but typically rather small relative change of measured tau uptake (say between 10\% and 20\%).
Our goal is to estimate the rate of amplification of misfolded tau-protein aggregates and the rate of fragmentation and migration thereof to distant parts of the brain. In particular, we study the sensitivity of such calibration to small relative changes in signal and tissue atrophy.
Our hypothesis is that an informative \emph{minimal} model can produce biomarkers, which can augment imaged-based approaches. Once calibrated, the current and future spatiotemporal spreading of tau can be quantified using our model.

\medskip\noindent\ipoint{Contributions.} We study the image-driven calibration of a biophysical model for spatiotemporal evolution of tau protein misfolding. Specifically, we investigate the effect of the time horizon between consecutive scans, and whether accounting for tissue atrophy affects the reconstruction of tau parameters. Lastly, we study how our method applies to real data and how imaging noise can be handled.
The novelty of our work can be summarized as follows:

\ipoint{1)}~We formulate and solve an inverse problem to estimate patient specific, characteristic growth parameters describing the spatiotemporal evolution of misfolded tau-protein throughout the brain based on longitudinal 3D tau-PET imaging of AD subjects.
We model tau progression as a reaction-diffusion-advection PDE, which accounts for tau propagation and takes into account observed atrophy. The atrophy is modeled using material transport with a velocity obtained from diffeomorphic image registration, and is one-way coupled to tau progression.

\ipoint{2)}~We investigate and demonstrate the ability to accurately estimate the model's growth parameters. We perform a sensitivity study with respect to the relative change in tau signal between scans, and the effect of tissue atrophy. For this experiment, we use synthetic data. Our results indicate good agreement for future prediction of tau uptake compared to the true data. 

\ipoint{3)}~We test our method on clinical tau-PET scans from the Alzheimer’s Disease Neuroimaging Initiative (ADNI)\footnote{\url{http://adni.loni.usc.edu/}} database, and study the sensitivity of our model to different subjects, and algorithmic hyper-parameters. To our knowledge, this is the first study of its kind. %

\medskip\noindent\ipoint{Related work.}
Biophysical modeling for AD started off with the advent of the prion-paradigm (misfolding chain-reaction) disease model. Models range from molecular level~\cite{Cohen:1994a,Jarrett:1993a}, to graph models~\cite{Abdelnour:2014,Raj:2012,Iturria-Medina:2017,Davis:2015}, and kinetic equations~\cite{Bertsch:2016}.
Inspired by the successful application of such models in computational oncology~\cite{yankeelov-miga13,alfonso2017biology,Hogea:2008b,Lipkova2019a}, we use an organ level PDE-model which reflects the most dominant evolution patterns of tau propagation. For AD, a similar model to the one we propose here has been recently proposed for prion-like diseases~\cite{Weickenmeier:2018,Weickenmeier:2019}. To the best of our knowledge, our work here is the first to study image-driven calibration of a biophysical model for tau propagation.

The remainder of the paper is structured as follows: \secref{sec:meth} discusses the physics-based model for the propagation of tau-protein, followed by the formulation and numerical solution of the inverse problem for image-driven calibration. In~\secref{sec:res}, we investigate the invertability of growth parameters, perform a sensitivity study, and apply our method to clinical data. We conclude the paper with a discussion in~\secref{sec:dis}.

\section{Methods}\label{sec:meth}

\subsection{Models and Materials}\label{sec:model-material}

\medskip\noindent\ipoint{Forward Model.}
To model the spatiotemporal evolution of misfolded tau-protein, we adopt the Fisher-Kolmogorov equation~\cite{Murray:1989a} coupled with an advection equation for material transport:
\begin{subequations}\label{eq:tu-state}
\begin{align}
    & &\p_t c + \idiv (c \vect{v}) - \kappa \D{D}c - \rho \D{R} c =  0, & &&\mbox{in}~\Omega_{\D{B}}\times (0,T], \label{eq:fwd} \\
    & &c(0) = c_0 & &&\mbox{in}~ \Omega_{\D{B}}, \label{eq:fwd-init}\\
    & &\partial_t \vect{m} + \igrad\vect{m} \vect{v}  =  \vect{0}, & &&\mbox{in}~ \Omega \times (0,T], \label{eq:adv-eq1}  \\
    & &\vect{m}(0) = \vect{m}_0 & &&\mbox{in}~\Omega. \label{eq:adv-eq2}
\end{align}
\end{subequations}
This simple model captures the basic dynamics of the problem: \bipa \item The growth and formation of tau aggregates, and \item their subsequent fragmentation and spatial propagation. \eipa
It further represents AD specific characteristics such as slow early stage progression with a rapid acceleration after symptom onset, and the inevitable progression of the disease: even a single corruptive protein will spread, and ultimately cause disease~\cite{Masel:99a}.

Our model follows~\cite{Weickenmeier:2018,Weickenmeier:2019}, but differs in a new image-driven transport term, that captures atrophy without a mechanical model.
In our case $c=c(\vect{x},t) \in [0,1]$ is the tau concentration with initial seeding $c_0$. Its evolution over time $t\in [0,T]$ in the three-dimensional brain domain $\Omega_{\D{B}} \subset \Omega = [0, 2\pi]^3$ is governed by the coefficients of the reaction and diffusion terms in~\eqref{eq:fwd}:
The nonlinearity $\D{R}c = \mat{\rho_{\vect{m}}}c(1-c)$ with growth rate $\rho$ provides a saturation term expressing the maximal concentration of toxic proteins.
$\mat{\rho}_{\vect{m}}$ is a spatially variable coefficient, depended on the underlying material properties $\vect{m} = \left( m_i(\vect{x},t) \right)_{i = W, \,G, \,F}$, a vector of voxel-wise probabilities for white matter $(W)$, gray matter $(G)$, and cerebrospinal fluid with ventricles $(F)$.
Spatial spreading of tau is driven by \emph{extracellular diffusion} and \emph{axonal transport}. This is modeled by the diffusion operator $\D{D}c = \idiv \mat{\kappa_{\vect{m}}}\igrad c$, where $\mat{\kappa_{\vect{m}}} = \mat{\kappa}_0\mat{I} + (\nf{\kappa_i}{\kappa} - 1) \mat{T}$ defines the inhomogeneous (anisotropic) diffusion tensor. $\mat{\kappa}_0$ captures the inhomogeneous diffusion based on the material properties $\vect{m}$, while $\mat{T}$ expresses preferential direction of diffusion along the axon bundle weighted by $\kappa_i$, and enables anisotropy.
Tau does not invade the cerebrospinal fluid, and spreading occurs primarily in white matter~\cite{Iaccarino:2018}.

We approximate no-flux boundary conditions $\partial \Omega_{\D{B}}$ via a penalty approach~\cite{Gholami:2016a}, and use periodic boundary conditions on $\partial \Omega$.

\medskip\noindent\ipoint{Atrophy Representation.}
Over time, tau aggregates disrupt cell function and ultimately cause cell death and tissue atrophy (thinning of white and gray matter)~\cite{La-Joie:2020}.
We represent tissue atrophy as image-driven transport equation~\eqref{eq:adv-eq1}--\eqref{eq:adv-eq2} of the spatiotemporal material properties $\vect{m}$.
As a result, tissue atrophy is one-way coupled to tau spreading via the definition of the spatially variable diffusion and reaction coefficients.

The velocity field $\vect{v}$ is found via large-deformation diffeomorphic image registration of longitudinal MRI data, and can be computed solving the inverse problem
\begin{equation}\label{eq:reg}
\min_{\vec{v}}
\frac{1}{2}\| \vec{m}(1) - \vec{m}_T \|_{L_2(\Omega)}^2 + \frac{\beta}{2} \mathcal{S}(\vect{v}) ~~~~\mbox{s.t.}~~ \mbox{\eqref{eq:adv-eq1}--\eqref{eq:adv-eq2}}.
\end{equation}
Here, $\vec{m}_0$ and $\vec{m}_T$ denote segmentations of MRI scans, corresponding to the acquisition times $t=0$ and $t=T$ of the tau-PET time-series.
The regularization term $\D{S}(\vect{v})$ is given as an $H^1$-seminorm $\int_{\Omega}\sum_{i-1}^3 |\igrad v^i(\vect{x}) |^2 \dx$. For the solution of~\eqref{eq:reg} we use the registration software \texttt{CLAIRE}~\cite{Mang2018b:CLAIRE}.
The proposed atrophy model is distinct from other approaches in the literature. In~\cite{Weickenmeier:2018,Weickenmeier:2019} the authors couple a reaction-diffusion equation to a mechanical deformation model for atrophy, based on nonlinear elasticity.

Solving~\eqref{eq:tu-state} defines the \emph{forward problem} $\D{F}(c_0, c_T, \vec{v}, \rho, \kappa) = 0$. Next, we discuss the image-driven calibration of this model.

\subsection{Model Calibration and Extraction of Tau-Spreading Characteristics}\label{sec:inverse-problem}
The model is personalized based on patient specific, tau-PET imaging data.
That is, we estimate biophysical growth parameters $\vect{g} = (\rho, \kappa)$ in~\eqref{eq:tu-state} based on measurements of tau uptake value ratios (SUVR) from longitudinal PET by solving a \ipoint{parameter estimation problem}
\begin{equation}\label{eq:inv-prob}
\min_{\vect{g} = (\rho, \kappa)}
\frac{1}{2}\|\mathcal{O}(c_T - d_T) \|_{L_2(\Omega)}^2 ~~\mbox{s.t.}~ \D{F}(c_0, c_T, \vec{v}, \rho, \kappa) = 0, ~\text{from}~\eqref{eq:tu-state},
\end{equation}
minimizing the discrepancy between predicted tau concentration $c_T=c(T)$ (based on the seeding $c(0)$) and the target data $d_T$.
The \ipoint{observation operator} $\mathcal{O}$ allows to specify a `valid` region (i.e., region of high signal confidence/relevance) of the tau-PET scan, which drives the parameter inversion. Although there has been success in reducing non-specific tracer binding in white matter, it remains to be problematic and white matter regions are disregarded when interpreting tau-PET scans~\cite{lyoo2018tau,Wang2019a}. Following this practice, we restrict the data term to gray matter only, and define $\mathcal{O}c \defeq \mathbb{1}_{[\vect{x}\in GM]}c$.
The seeding concentration $c(0)$ is defined by the data $d_0$ from the first time point $d_0 =: c_0$.
As a proof of concept, we only consider isotropic diffusion for the model calibration. The extension to anisotropy is straightforward and will be realized in future work.
Before, discussing the numerical scheme for the solution of~\eqref{eq:inv-prob}, we want to discuss some characteristics of the parameter inversion using a simplified analytical model in one dimension.

\subsection{Parameter Inversion Analysis for Linearized Model}
Although our model has only two free parameters, the inversion problem~\eqref{eq:inv-prob} may be ill-conditioned and suffer from strong sensitivity to perturbations.
To demonstrate some of the inherent instabilities, we study the analytic solution for a simplified version of the forward model~\eqref{eq:tu-state}. We linearize around $c(x,t) = 0$, take $\vect{v}=0$, and consider constant coefficients in a one-dimensional domain $\omega = [0,\pi]$  with periodic boundary conditions:
\begin{subequations}\label{eq:linearized-state}
\begin{flalign}
    \partial_t c - \kappa \partial_{xx} c - \rho c &= 0 ~~~\mbox{in}~\omega \times (0,T], \\
    c(0) &= c_0 ~~\mbox{in}~ \omega, ~~~~~~\partial_xc(0) = \partial_xc(\pi) = 0.
\end{flalign}
\end{subequations}
The BVP \eqref{eq:linearized-state} can be solved analytically by separation of variables, yielding
\begin{equation}\label{eq:linear-solution}
    c(x,t) = \sum_{n=0}^{\infty} \hat{c}_n\exp\left((\rho - n^2\kappa)t\right)\cos(nx),
\end{equation}
where $\hat{c}_n$ are the spectral cosine coefficients of the initial condition $c_0$. Assuming band-limited data $d_T = \sum_{n=0}^N\hat{d}_n\cos(nx)$, the parameter inversion problem for $\rho$, $\kappa$ reads
\begin{equation}\label{eq:linear-inverse-problem}
    \mbox{min}_{\rho,\kappa} = \frac{1}{2}\sum_{n=0}^{N} \left(\hat{c}_n\exp\left((\rho - n^2\kappa)T\right) - \hat{d}_n\right)^2.
\end{equation}
It is easy to see that a minimum is found if
\begin{equation}\label{eq:minimum-cond}
    \hat{c}_n\exp\left((\rho - n^2\kappa)T\right) = \hat{d}_n ~~\mbox{or}~~ \frac{1}{T}\log{\frac{\hat{c}_n}{\hat{d}_n}} = -\rho + n^2\kappa.
\end{equation}
From \eqref{eq:minimum-cond} we observe, that any numerical errors, noise, or model errors are expected to be amplified when the time horizon decreases.

\subsection{Numerical Scheme}\label{sec:methods-numerics}
\medskip\noindent\ipoint{Numerical Solution of the Forward Problem.}
We employ a pseudo-spectral Fourier approach on a regular grid for spatial discretization.
That means all spatial differential operators are computed via a 3D fast Fourier Transform~\cite{Gholami:2017a}. For numerical solution of the forward problem, we employ a first-order operator-splitting method to split the tau progression equation~\eqref{eq:fwd} into a reaction, diffusion, and advection part. For the diffusion split, we use an implicit Crank-Nicholson method, and solve the linear system with a preconditioned matrix-free CG method. The reaction sub-steps are solved analytically.
The \emph{hyperbolic transport equations} are solved using an unconditionally stable semi-Lagrangian time-stepping scheme to avoid stability issues and small, CFL restricted time-steps~\cite{Mang:2016c,Mang:2017b}.
This method requires evaluations of the space-time fields at off-grid locations, defined by the characteristic associated with $\vect{v}$. We compute off-grid evaluations using a cubic Lagrange polynomial interpolation model.
We use $n_t=100$ time steps of size $\Delta t = 0.01$ for the time integration.

\medskip\noindent\ipoint{Numerical Optimization.}
The optimization problem~\eqref{eq:inv-prob} is solved using a bound constrained, limited memory BFGS quasi-Newton solver globalized with Armijo line-search. The gradient is computed using a first-order accurate finite difference scheme with $\vec{h}=\sqrt{\epsilon_{mach}}\vec{g}$, for $\vec{g} = (\rho, \kappa)^T$. We terminate the optimization after a gradient reduction of $4$ orders of magnitude (relative to the initial guess).
To keep the optimizer within feasible bounds, and prevent bad local minima, we define bound constraints $\kappa_{min}=\num{1E-4}$, and $\kappa_{max}=1$, $\rho_{min}=0.1$, and $\rho_{max}=15$.\footnote{The lower bound for $\kappa_{i}$ is determined by the smallest recoverable diffusivity considering a resolution of $256^3$.}

\subsection{Workflow Summary}\label{sec:preproc}
For the evaluation on clinical data, we use seven AD subjects from the ADNI database as outlined in~\tabref{tab:ClinPET}.
Imaging data is processed using FSL~\cite{Jenkinson:FSL} by applying the following steps:
\bipa
\item longitudinal MRI scans are rigidly registered to the first time point using FLIRT~\cite{Jenkinson:FLIRT}; \item longitudinal tau-PET images are rigidly registered to the first time point MRI; \item aligned images are skull-stripped using BET~\cite{Jenkinson:BET}; \item registered brain masks are applied in PET space to obtain the skull-stripped PET image; \item skull-stripped images are segmented using FAST~\cite{Zhang:2001}; \item PET images are individually normalized with average tau uptake value in the cerebellum.
\eipa For the last step, the cerebellum is extracted by registering the ADNI subject image to labeled atlas.

\section{Results}\label{sec:res}

We examine the quality of biophysical model personalization, and the accuracy of subsequent prediction of tau spreading.
We ask the following two questions: \bipa \item How does the solution of~\eqref{eq:inv-prob} depend on the time horizon $T$ and the effects of tissue atrophy (modeled via material transport)? And \item how does our method perform on clinical tau-PET scans? \eipa

\subsection{Accuracy of Model Calibration and Tau Forecast Using Synthetic Data}\label{sec:synthetic-results}
\begin{landscape}
\begin{figure}
  \centering\hspace{-1cm}
  \centering
    \includegraphics[height=\textwidth]{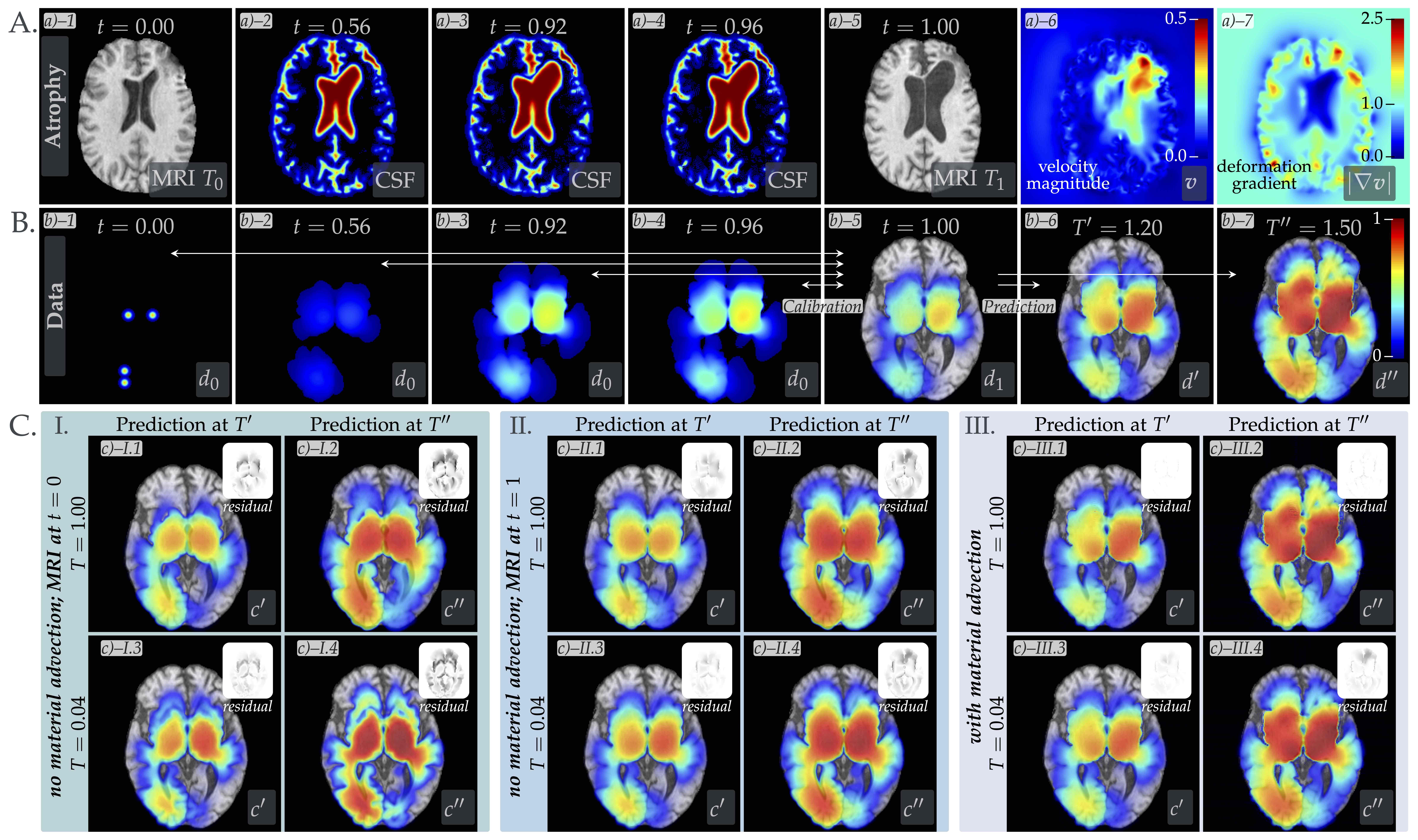}
    \caption{\emph{Qualitative results for model calibration and tau evolution forecasting using synthetic data.}
    Panel A illustrates structural differences induced by tissue atrophy: a)--1 and a)--5 show original  MRI scans of an AD subject at time points $T_0=0$ and $T_1=1$; a)--2 through a)--4 show the change of the CSF structure at different time points due to atrophy; a)--5 and 6 show the magnitude and deformation gradient of the velocity, obtained from registration of  MRI (a value of $|\igrad\vect{v}| < 1$ indicates contraction, a value above 1 expansion; the registration is performed from the second to the first snapshot).
    Panel B shows synthetic tau data at different time points used for calibration and tau distribution at forecast time points $T^{\prime}$ and $T^{\prime\prime}$
    (cf.\,\tabref{tab:syn-results-sensitivity-reg-timeframe}).
    Panel C shows the forecast distribution $c^{\prime}$ and $c^{\prime\prime}$ of tau at time points $T^{\prime}$ and $T^{\prime\prime}$ based on calibration for the three scenarios I-III from~\tabref{tab:syn-results-sensitivity-reg-timeframe} and two different time horizons, along with the relative mismatch to the ground truth (dark black indicates high error).
    }
    \label{fig:rho-kappa-inv:SynPET_2}
\end{figure}
\end{landscape}

We study the sensitivity of model calibration and prediction to
\bipa
\item small relative change in tau signal (varying time frames of image acquisition), and
\item the effect of tissue atrophy
\eipa
for a synthetic setup. \figref{fig:rho-kappa-inv:SynPET_2} gives an illustration of tau propagation and the extend of atrophy.

\medskip\noindent\ipoint{Synthetic Data.}
As baseline brain anatomy, we use (segmentations of) longitudinal  MRI scans of an AD subject with clearly visible tissue atrophy (cf.\,panel A in~\figref{fig:rho-kappa-inv:SynPET_2}). The velocity $\vect{v}$, used to couple the effect of tissue atrophy to tau spreading via the advection term in \eqref{eq:tu-state}, is obtained using diffeomorphic image registration between segmentations of the MRI of the first and second scan. We generated synthetic tau data using our reaction-diffusion-advection model for tau evolution based on the tissue segmentation of the first time point, and ground truth parameters (proliferation rate of $\rho=8$, and migration rate of $\kappa=0.18$; nondimensionalized). Synthetic data was generated using twice as many discretization points in time as compared to the inversion. An illustration is given in panel B of \figref{fig:rho-kappa-inv:SynPET_2}.
The primary purpose of these synthetic experiments is to set up a simple problem to study the effects of tissue atrophy and small relative change in signal. Thus, the synthetic data is observed everywhere and observations are noise free.

We consider three scenarios:
\begin{itemize}\setlength\itemsep{0.0em}
\item[\phantom{II}I.] Disregard tissue atrophy and define material properties for tau-propagation based on structural MRI at first time point $T_0$;
\item[\phantom{I}II.] Repeat the same setup but now use the structural MRI at time point $T_1$;
\item[III.] Account for tissue atrophy via material transport, governed by a velocity field obtained from image registration of (segmented) MRI.
\end{itemize}
For each scenario, the time horizon between image acquisition of consecutive scans is varied (the first snapshot $d_0$ is taken at different times $T_0=0.0, 0.56, \ldots, 0.99$), with a relative change in tau uptake signal ranging from 99\% to 3\%. Since image acquisition time intervals are rather short in practice (8-15 months), studying the sensitivity of the calibration to small relative changes in the signal, is of great interest. Results are given in~\tabref{tab:syn-results-sensitivity-reg-timeframe}.

\medskip\noindent\ipoint{Performance Metrics.}
We report relative errors $\epsilon_{\iota} = \nf{\iota_{rec}}{\iota^{\star}}$, $\iota \in \{\rho, \kappa\}$ in the reconstruction of the growth parameters. We further report the relative forecast errors $\mu_{t} = \nf{\|c(t) - d_t\|_2}{\|d_t\|_2}$ for predicted tau concentration at future time points $t\in\{T^{\prime}, T^{\prime\prime}\}$. All numerical experiments are indicated with a unique identification number \#XY (first column of tables).

\medskip\noindent\ipoint{Sensitivity to Small Relative Change in tau Signal.}
Runs \#17--\#24 (Scenario III) in  \tabref{tab:syn-results-sensitivity-reg-timeframe} show the calibration results and subsequent tau forecast for varying acquisition times of the first scan $d_0$. We observe a slight deterioration in reconstruction accuracy of the true model parameters for smaller time horizons between scans. For particularly small relative change in tau uptake, we see larger errors and a trend of under-estimating the growth parameters. Up to 10\% relative change in tau SUV, relative errors, however, are still small with 5\% and below for $\rho$, and up to 20\% for $\kappa$.
Similarly the accuracy of tau forecast deteriorates marginally, but remains low with errors of 3-13\% for $T^{\prime}$, and 5-26\% for $T^{\prime\prime}$ (see also panel C in~\figref{fig:rho-kappa-inv:SynPET_2}).

\begin{table}[htb]
  \centering\small\setlength\tabcolsep{4pt}
  \caption{\emph{Model calibration and tau evolution forecasting for synthetic data.} We study the sensitivity with respect to 1) the time horizon $T$ between consecutive scans, and 2) tissue atrophy (one-way coupled to tau spreading via material transport). For 1), we consider varying image acquisition time points $T_0$ for $d_0$ with a relative change ($\Delta$ SUV) in tau signal (between scans $d_0$ and $d_1$) ranging from 99\% to 3\%. For 2), we consider three different scenarios: I. Disregard tissue atrophy and use structural MRI of $T_0$ as material properties for tau-propagation; II. Like before but use MRI of $T_1$; III. Account for tissue atrophy via material transport, governed by a velocity field found from image registration of (segmented) $T_1$ and $T_0$ MRI.
  We report relative errors $\epsilon_{\rho}$ and $\epsilon_{\kappa}$ for the inversion parameters (true values are $\rho^{\star}=8$, $\kappa^{\star}=0.18$). $\mu_{T_1}$ denotes the relative data misfit in the inversion; $\mu_{T^{\prime}}$ and $\mu_{T^{\prime\prime}}$ denote forecast errors at future times $T^{\prime}=1.2$, and $T^{\prime\prime}=1.5$.
  \label{tab:syn-results-sensitivity-reg-timeframe}
  }
  \resizebox{\textwidth}{!}{
\begin{tabular}{lllll|llLLLLL}
    \toprule
ID & & $T_0$ & $T_1$ & $\Delta$ SUV & $\rho$ & $\kappa$ & $\epsilon_{\rho}$ & $\epsilon_{\kappa}$ & $\mu_{T_1}$ & $\mu_{T^{\prime}}$ & $\mu_{T^{\prime\prime}}$ \\
\midrule
$\runid$ &
\multirow{8}{*}{\rotatebox{90}{\scriptsize\textbf{\textit{I. $T_0$ MRI}}}}
& 0.00 & 1.00  & 99\% & 7.77 & \num{1.703830e-01}& \num{2.905875e-02} & \num{5.342778e-02} & \num{3.374800e-01} & \num{3.897550e-01} & \num{4.422460e-01}   \\
$\runid$ & & 0.56 & 1.00  & 80\%  & 7.89 & \num{1.852090e-01}& \num{1.367625e-02} & \num{2.893889e-02} & \num{3.034890e-01} & \num{3.696890e-01} & \num{4.292840e-01}   \\
$\runid$ & & 0.72 & 1.00  & 60\%  & 7.70 & \num{1.523760e-01}& \num{3.703250e-02} & \num{1.534667e-01} & \num{2.556090e-01} & \num{3.427360e-01} & \num{4.175330e-01}   \\
$\runid$ & & 0.83 & 1.00  & 40\%  & 7.21 & \num{9.364880e-02}& \num{9.925375e-02} & \num{4.797289e-01} & \num{1.905430e-01} & \num{3.125620e-01} & \num{4.173920e-01}   \\
$\runid$ & & 0.92 & 1.00  & 20\%  & 6.48 & \num{4.120840e-02}& \num{1.895475e-01} & \num{7.710644e-01} & \num{1.124470e-01} & \num{2.801030e-01} & \num{4.270200e-01}   \\
$\runid$ & & 0.96 & 1.00  & 10\%  & 5.96 & \num{1.494330e-02}& \num{2.548425e-01} & \num{9.169817e-01} & \num{6.030390e-02} & \num{2.626310e-01} & \num{4.387130e-01}   \\
$\runid$ & & 0.98 & 1.00  & 5\%   & 5.78 & \num{6.920230e-03}& \num{2.770450e-01} & \num{9.615543e-01} & \num{3.089030e-02} & \num{2.515650e-01} & \num{4.427930e-01}   \\
$\runid$ & & 0.99 & 1.00  & 3\%   & 5.73 & \num{4.520080e-03}& \num{2.839250e-01} & \num{9.748884e-01} & \num{1.559450e-02} & \num{2.445290e-01} & \num{4.432530e-01}   \\
\midrule
$\runid$ &
\multirow{8}{*}{\rotatebox{90}{\scriptsize\textbf{\textit{II. $T_1$ MRI}}}}
& 0.00 & 1.00  & 99\% & 7.53 & \num{1.588220e-01}& \num{5.929500e-02} & \num{1.176556e-01} & \num{2.490390e-01} & \num{2.660540e-01} & \num{3.062370e-01}   \\
$\runid$ & & 0.56 & 1.00  & 80\%  & 7.71 & \num{1.648790e-01}& \num{3.569000e-02} & \num{8.400556e-02} & \num{1.837620e-01} & \num{2.219050e-01} & \num{2.772380e-01}   \\
$\runid$ & & 0.72 & 1.00  & 60\%  & 7.70 & \num{1.555860e-01}& \num{3.720625e-02} & \num{1.356333e-01} & \num{1.481570e-01} & \num{2.010840e-01} & \num{2.667910e-01}   \\
$\runid$ & & 0.83 & 1.00  & 40\%  & 7.58 & \num{1.366370e-01}& \num{5.191125e-02} & \num{2.409056e-01} & \num{1.135070e-01} & \num{1.830010e-01} & \num{2.597460e-01}   \\
$\runid$ & & 0.92 & 1.00  & 20\%  & 7.28 & \num{1.068850e-01}& \num{9.001375e-02} & \num{4.061944e-01} & \num{7.350680e-02} & \num{1.664800e-01} & \num{2.599070e-01}   \\
$\runid$ & & 0.96 & 1.00  & 10\%  & 6.83 & \num{7.458450e-02}& \num{1.464000e-01} & \num{5.856417e-01} & \num{4.387150e-02} & \num{1.611420e-01} & \num{2.750100e-01}   \\
$\runid$ & & 0.98 & 1.00  & 5\%   & 6.31 & \num{4.143880e-02}& \num{2.115775e-01} & \num{7.697844e-01} & \num{2.422540e-02} & \num{1.688220e-01} & \num{3.060040e-01}   \\
$\runid$ & & 0.99 & 1.00  & 3\%   & 6.02 & \num{2.355370e-02}& \num{2.479012e-01} & \num{8.691461e-01} & \num{1.258370e-02} & \num{1.772650e-01} & \num{3.289820e-01}   \\
\midrule
$\runid$ &
\multirow{8}{*}{\rotatebox{90}{\scriptsize\textbf{\textit{III. Adv., $T_0$ MRI}}}}
& 0.00 & 1.00  & 99\% & 7.90 & \num{1.800850e-01}& \num{1.310500e-02} & \num{4.722222e-04} & \num{2.379750e-02} & \num{3.343690e-02} & \num{5.230390e-02}   \\
$\runid$ & & 0.56 & 1.00  & 80\%  & 7.90 & \num{1.830770e-01}& \num{1.198000e-02} & \num{1.709444e-02} & \num{9.744980e-02} & \num{1.016240e-01} & \num{1.112870e-01}   \\
$\runid$ & & 0.72 & 1.00  & 60\%  & 7.89 & \num{1.807570e-01}& \num{1.374750e-02} & \num{4.205556e-03} & \num{9.214580e-02} & \num{1.089900e-01} & \num{1.268580e-01}   \\
$\runid$ & & 0.83 & 1.00  & 40\%  & 7.83 & \num{1.716360e-01}& \num{2.158875e-02} & \num{4.646667e-02} & \num{7.300130e-02} & \num{1.061330e-01} & \num{1.334900e-01}   \\
$\runid$ & & 0.92 & 1.00  & 20\%  & 7.70 & \num{1.574190e-01}& \num{3.708375e-02} & \num{1.254500e-01} & \num{4.605880e-02} & \num{9.904710e-02} & \num{1.381360e-01}   \\
$\runid$ & & 0.96 & 1.00  & 10\%  & 7.56 & \num{1.430790e-01}& \num{5.450500e-02} & \num{2.051167e-01} & \num{2.657260e-02} & \num{9.423950e-02} & \num{1.433210e-01}   \\
$\runid$ & & 0.98 & 1.00  & 5\%   & 6.66 & \num{7.219130e-02}& \num{1.680300e-01} & \num{5.989372e-01} & \num{1.763950e-02} & \num{1.219130e-01} & \num{2.186460e-01}   \\
$\runid$ & & 0.99 & 1.00  & 3\%   & 5.97 & \num{6.729410e-02}& \num{2.535825e-01} & \num{6.261439e-01} & \num{9.894320e-03} & \num{1.371280e-01} & \num{2.642340e-01}   \\
\bottomrule
\end{tabular}}
\end{table}

\medskip\noindent\ipoint{Sensitivity to Effects from Tissue Atrophy (Material Transport).}
Next, we study the sensitivity of the model calibration due to atrophy-induced structural changes, that is the distribution of CSF, gray and white matter.
From \tabref{tab:syn-results-sensitivity-reg-timeframe} we observe that the effect of tissue atrophy is not negligible for the calibration of our model: Disregarding it misses the effects of cortical thinning, and this results in quite large errors for both scenarios I and II, compared to the full model in scenario III. Most noticeably, smaller relative change of tau SUV between scans results in a loss of accuracy for both the calibrated growth parameters and the forecast of tau concentration. This accuracy loss is more severe, if the advection of material properties, according to the cortical thinning, is disregarded. The differences in the forecast of tau distribution at future time points can be seen in~\figref{fig:rho-kappa-inv:SynPET_2} (panel C) for the different scenarios.
We conclude that the model is sensitive in the presence of large atrophy and the coupling of tissue atrophy, captured by the transport velocity $\vect{v}$, is important for reconstructing the correct parameters. Of course, this sensitivity depends on the extend of observable atrophy in the longitudinal MRI scans. If the atrophy is negligible, then the advection term in the tau propagation term will be almost zero.

\begin{table}
  \centering\small\setlength\tabcolsep{4pt}
  \caption{\ipoint{Model Calibration for ADNI Clinical Data.}
  We report the calibrated growth parameters $\rho$ (in [$\times 10^{-3}/$day]) and $\kappa$ (in [$\times 10^{-3}$mm$^2$/day]) for seven AD patients, from the ADNI database.
  We preprocess the already smoothed PET tau values by subtracting $f_b\times$ the mean signal in cerebellum from the normalized tau-PET scan. We run the calibration for a factor of $f_b=1.6$ as suggested in the tauvid protocol (highlighted in red), but also examine the sensitivity of the model calibration with respect to this parameter.
  Patient demographics as well as Mini-Mental State Exam (MMSE) scores and patient group (Alzheimer's disease (AD), early cognitive impairment (EMCI), late cognitive impairment (LMCI), cognitively normal (CN)) are given.
  The time $T$ between scans is given in days. If available, we register  MRIs of longitudinal snapshot, and account for atrophy (via material advection) in the personalization step. Entries with ``--'' indicate inability to run our model because one or both of $d_0$ and $d_1$ are zero.
  }
  \label{tab:ClinPET}
  \resizebox{\textwidth}{!}{
  \begin{tabular}{lLlllllll}
  \toprule
  \rowcolor{colorA!20}
  \textbf{$f_b$} & \textbf{ID} & \textbf{022\_S\_6013} & \textbf{023\_S\_1190} & \textbf{127\_S\_4301} & \textbf{127\_S\_2234} & \textbf{035\_S\_4114} & \textbf{033\_S\_4179} & \textbf{941\_S\_4036} \\
  \midrule
  & Sex      & F     & F     & M    & F    & F    & M    & M    \\
  & Age      & 60.0  & 83.8  & 76.6 & 64.3 & 57.1 & 85.1 & 74.4 \\
  & MMSE     & N/A   & 28    & 30   & 29   & 27   & 27   & 28   \\
  & Group    & AD    & CN    & EMCI & EMCI & LMCI & CN   & EMCI \\
  \midrule
  & $T\,[d]$ & 401   & 476   & 477  & 660  & 426  & 376  & 464  \\
  & Atrophy  & yes   & yes   & no   & no   & yes  & no   & yes  \\
\midrule
\multirow{3}{*}{\rotatebox{90}{\textbf{1.0x}}}
  &  $\rho$  & \num{0.00}& \num{0.41}& \num{0.00}& \num{0.00}& \num{1.95}& \num{0.62}& \num{0.00}   \\
  &  $\kappa$ & \num{3.796960e-04}& \num{4.548930e-03}& \num{2.770770e-03}& \num{1.707640e-01}& \num{4.898680e-03}& \num{1.590780e-03}& \num{2.072250e-02}   \\
  &  $\mu$    & \num{4.149960e-01}& \num{7.580970e-01}& \num{6.631600e-01}& \num{8.993910e-01}& \num{6.658980e-01}& \num{6.583430e-01}& \num{7.758190e-01}   \\
\midrule
\multirow{3}{*}{\rotatebox{90}{\textbf{1.2x}}}
  &  $\rho$ & \num{0.04}& \num{0.69}& \num{0.00}& \num{0.91}& \num{3.15}& \num{1.58}& \num{0.14}   \\
  &  $\kappa$  & \num{3.350590e-04}& \num{3.983120e-03}& \num{3.245130e-03}& \num{1.670940e-01}& \num{8.128570e-03}& \num{2.065240e-03}& \num{1.594500e-02}   \\
  &  $\mu$ & \num{4.307920e-01}& \num{8.281310e-01}& \num{6.781590e-01}& \num{1.026960e+00}& \num{7.412240e-01}& \num{7.200520e-01}& \num{8.936590e-01}   \\
\midrule
\multirow{3}{*}{\rotatebox{90}{\textbf{1.4x}}}
  &  $\rho$ & \num{0.00}& \num{1.19}& \num{0.00}& \num{2.95}& \num{4.09}& \num{2.74}& \num{1.50}   \\
  &  $\kappa$ & \num{1.000000e-04}& \num{2.413610e-03}& \num{1.830200e-03}& \num{1.600580e-01}& \num{6.363420e-03}& \num{3.767820e-03}& \num{7.371710e-03}   \\
  &  $\mu$ & \num{4.698560e-01}& \num{8.018760e-01}& \num{5.986620e-01}& \num{1.052010e+00}& \num{7.881430e-01}& \num{7.506220e-01}& \num{8.865280e-01}   \\
\midrule
\multirow{3}{*}{\rotatebox{90}{\textbf{1.65$\times$}}}
  &  \cellcolor{red!20} $\rho$ & \num{3.24}& \num{1.75}& \num{0.00}& \num{3.30}& \num{4.93}& \num{4.97}& \num{3.45}   \\
  &  \cellcolor{red!20}  $\kappa$ & \num{1.937360e-03}& \num{3.236500e-03}& \num{7.923000e-04}& \num{8.900000e-01}& \num{3.758940e-03}& \num{8.386240e-03}& \num{3.321620e-03}   \\
  &  \cellcolor{red!20}  $\mu$ & \num{5.982070e-01}& \num{8.310760e-01}& \num{5.740260e-01}& \num{1.016150e+00}& \num{8.223700e-01}& \num{7.864160e-01}& \num{9.550130e-01}   \\
\midrule
\multirow{3}{*}{\rotatebox{90}{\textbf{1.8x}}}
  &  $\rho$ & \num{5.24}& \num{2.07}& \num{0.00}& -- & \num{5.47}& \num{6.88}& \num{4.00}   \\
  &  $\kappa$ & \num{2.437410e-03}& \num{4.876290e-03}& \num{5.697720e-04}& -- & \num{3.184880e-03}& \num{1.282300e-02}& \num{1.000000e-04}   \\
  &  $\mu$ & \num{6.801210e-01}& \num{8.877430e-01}& \num{5.833250e-01}& -- & \num{8.423150e-01}& \num{8.008640e-01}& \num{9.909770e-01}   \\
\midrule
\multirow{3}{*}{\rotatebox{90}{\textbf{2.0x}}}
  &  $\rho$  & \num{6.87}& \num{2.81}& \num{0.00}& -- & \num{6.39}& \num{7.97}& --   \\
  &  $\kappa$ & \num{1.510520e-03}& \num{1.00e-02}& \num{3.383570e-04}&  -- & \num{3.059260e-03}& \num{5.848770e-03}& --  \\
  &  $\mu$ & \num{9.272730e-01}& \num{9.589230e-01}& \num{6.017540e-01}& -- & \num{8.768500e-01}& \num{8.689610e-01}& --   \\
\bottomrule
\end{tabular}}
\end{table}

\subsection{Accuracy and Model Calibration on the ADNI Clinical Data}\label{sec:results-real-data}

\medskip\noindent\ipoint{Clinical Data.}
We use tau-PET scans from the ADNI database for calibration.
The ANDI-tau study is ongoing and the dataset is not yet complete. For this exploratory study, we selected cases with clear longitudinal differences in the tau signal and at least one good quality MRI. Subjects with only one tau scan or non-increasing tau uptake were excluded. Patient demographics are given in~\tabref{tab:ClinPET}. PET scans are acquired one to two years apart; the number of days between image acquisition is given by $T$. Not all cases have  MRI for the second time point; we do not account for atrophy changes for these subjects.

Raw ADNI imaging data is processed as described in~\secref{sec:preproc}. PET imaging typically features a low signal-to-noise ratio, which complicates the calibration process. Furthermore the ${}^{18}$F (tauvid) agent exhibits non-specific background activity of healthy tissue, and its  dynamic range is at an offset. The tauvid imaging protocol\footnote{\url{http://pi.lilly.com/us/tauvid-uspi.pdf}} suggests to correct for this ambient signal by subtraction of a constant-intensity `all black` image. The offset threshold was determined as  1.65$\times$ the mean tracer signal within the cerebellum ($\overline{CE}$).
Similarly, for our simulations, we preprocess all scans by subtracting a constant background intensity (multiple of the mean tracer signal in cerebellum).
As described in~\secref{sec:inverse-problem}, PET activity in white matter or outside the brain is non-specific and usually disregarded when interpreting tau-PET data. Thus, we fit our model to match target tau-PET signal in gray matter only. While white matter regions are not penalized in the data term of \eqref{eq:inv-prob}, tau protein may still spread along white matter fiber tracts. We assume a 100-fold higher migration rate in white matter, compared to gray matter.

Since the factor of 1.65 for the background subtraction seems somewhat arbitrary, we also study the sensitivity of the calibration (and reconstruction) with respect to this parameter.

\begin{landscape}
\begin{figure}
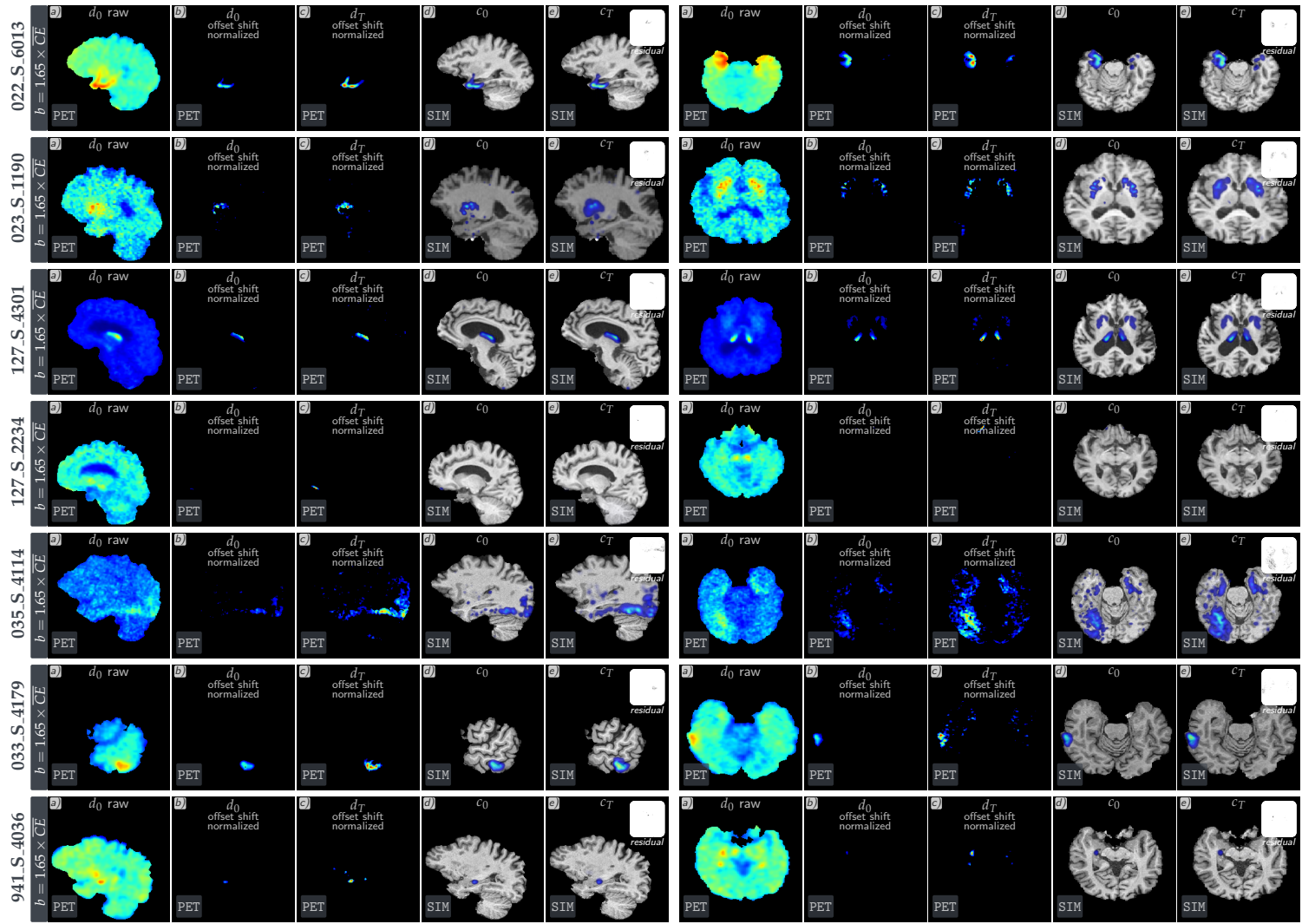

  \hspace{-0.8cm}
  \begin{tikzpicture}[
        every node/.style={anchor=south west,inner sep=0pt},
        x=1mm, y=1mm,
      ]
     \node[color=anthrazit] at (-0.3cm, 0.3cm) {\small \rotatebox{90}{ \textbf{022\_S\_6013}} };
     \node (fig1) at (0,0)
       {\includegraphics[scale=0.68]{figs/adni/offset_shift_tol1E-2/\detokenize{adni_subj_022_S_6013_line_bgcorr-1.65CE}.pdf}};
     \begin{scope}[shift={(0.0, -2.0cm)}]
     \node[color=anthrazit] at (-0.3cm, 0.3cm) {\small \rotatebox{90}{ \textbf{023\_S\_1190}} };
     \node (fig3) at (0,0)
       {\includegraphics[scale=0.68]{figs/adni/offset_shift_tol1E-2/\detokenize{adni_subj_023_S_1190_line_bgcorr-1.65CE}.pdf}};
     \begin{scope}[shift={(0.0, -2.0cm)}]
     \node[color=anthrazit] at (-0.3cm, 0.3cm) {\small \rotatebox{90}{ \textbf{127\_S\_4301}} };
     \node (fig3) at (0,0)
       {\includegraphics[scale=0.68]{figs/adni/offset_shift_tol1E-2/\detokenize{adni_subj_127_S_4301_line_bgcorr-1.65CE}.pdf}};
     \begin{scope}[shift={(0.0, -2.0cm)}]
     \node[color=anthrazit] at (-0.3cm, 0.3cm) {\small \rotatebox{90}{ \textbf{127\_S\_2234}} };
     \node (fig2) at (0,0)
       {\includegraphics[scale=0.68]{figs/adni/offset_shift_tol1E-2/\detokenize{adni_subj_127_S_2234_line_bgcorr-1.65CE}.pdf}};
     \begin{scope}[shift={(0.0, -2.0cm)}]
     \node[color=anthrazit] at (-0.3cm, 0.3cm) {\small \rotatebox{90}{ \textbf{035\_S\_4114}} };
     \node (fig3) at (0,0)
       {\includegraphics[scale=0.68]{figs/adni/offset_shift_tol1E-2/\detokenize{adni_subj_035_S_4114_line_bgcorr-1.65CE}.pdf}};
     \begin{scope}[shift={(0.0, -2.cm)}]
     \node[color=anthrazit] at (-0.3cm, 0.3cm) {\small \rotatebox{90}{ \textbf{033\_S\_4179}} };
     \node (fig3) at (0,0)
       {\includegraphics[scale=0.68]{figs/adni/offset_shift_tol1E-2/\detokenize{adni_subj_033_S_4179_line_bgcorr-1.65CE}.pdf}};
     \begin{scope}[shift={(0.0, -2.cm)}]
     \node[color=anthrazit] at (-0.3cm, 0.3cm) {\small \rotatebox{90}{ \textbf{941\_S\_4036}} };
     \node (fig1) at (0,0)
       {\includegraphics[scale=0.68]{figs/adni/offset_shift_tol1E-2/\detokenize{adni_subj_941_S_4036_line_bgcorr-1.65CE}.pdf}};

     \end{scope}
     \end{scope}
     \end{scope}
     \end{scope}
     \end{scope}
     \end{scope}
   \end{tikzpicture}
  \caption{\ipoint{Model personalization using clinical data for seven ADNI subjects with an offset of $1.65\times\overline{CE}$ for the background subtraction.}
  Shown are sagittal and axial cuts of tau-PET data and reconstruction.
  Panel a) shows axial and sagittal cuts of the raw tau-PET (for time point $T_0$); panels b) and c) show the background corrected tau scans (for time point $T_0$ and $T_1$, respectively); panels d) and e) show the initial tau seeding $c(0)$, and reconstruction $c_T$, respectively.
  }
  \label{fig:ClinPET1}
\end{figure}
\end{landscape}

\medskip\noindent\ipoint{Calibration Results.}
Quantitative and qualitative calibration results for varying background offsets are given in~\tabref{tab:ClinPET} and Figs.\;\ref{fig:ClinPET1}--\ref{fig:ClinPET6}.
Looking at the suggested factor of 1.65$\times$ the mean signal in cerebellum (highlighted in red in~\tabref{tab:ClinPET}), we observe that the estimated growth parameters show significant variability across different subjects.
For example, we obtain values for $\rho$ between close to zero and 4.97, and $\kappa$ ranges between $\num{8.9E-1}$ and $\num{7.92E-4}$. This demonstrates that our model is sensitive to patient-specific information and provides an indication that the methodology could be clinically useful for differential diagnosis.

Second, the model calibration is quite sensitive to the offset chosen for the background subtraction. We observe large variation for the estimated growth parameters $\rho$ and $\kappa$ across different offsets for the different subjects. \figref{fig:ClinPET1} shows the calibration data and reconstruction for an offset of $1.65\times\overline{CE}$ for the background subtraction: panel a) shows axial and sagittal cuts of the raw tau-PET (for time point $T_0$); panels b) and c) show the background corrected tau scans (for time point $T_0$ and $T_1$, respectively); panels d) and e) show the initial tau seeding $c(0)$, and reconstruction $c_T$, respectively. Figures~\ref{fig:ClinPET2}--\ref{fig:ClinPET6} show calibration data and reconstruction for varying offsets used for background subtraction.
For subject 035\_S\_4114, e.g., the estimated proliferation rate $\rho$ varies from 1.95 to 6.39 across different offsets for the background subtraction, while the estimation of the migration rate $\kappa$ is less sensitive. A similar trend can be seen for the other subjects.

\section{Discussion}\label{sec:dis}

The pronounced sensitivity of the calibration parameters to the offset for the background subtraction (but also to the overall imaging preprocessing, including normalization) is partially due to the simplistic nature of our model, and its strong dependency on the initial tau seeding.  The latter largely influences growth pattern and shape; since it is directly taken from the $T_0$ tau-PET scan, the offset for background subtraction enters the model in a direct and sensitive way. Reviewing Figs.\;\ref{fig:ClinPET2}--\ref{fig:ClinPET6}, we see that the actual region of interest (ROI) with significant change in tau signal is typically very localized. Choosing the offset too small, leaves residuals of background signal distributed throughout the brain, which, ultimately prevents the model from reconstructing (small and localized) high intensity peaks (since the larger areas of low intensity signal weigh higher in the $L_2$ misfit of the objective function). Subtracting a too large `background intensity` results in loss of information. Another problem arises when ROIs which showed medium to low signal intensity in the $T_0$ scan, appear with high signal intensity in the $T_1$ scan. This is a quite common scenario, yet poses challenges on our model since the initial tau seeding likely does not have support in such regions---either because they are indistinguishable from the background signal (and removed), or because indeed the ROI was not invaded by tau at $T_0$.

To capture such effects and overcome the shortcomings described above, a proper noise model must be incorporated, and parameter estimation needs to be performed in a Bayesian manner. We will target this approach in a follow up study. Another important model improvement is to include anisotropic diffusion, especially for data with longer time horizon $T$. Also in our model, the atrophy is computed from image registration and it is only one-way coupled to tau. In our future work, we will evaluate such models~\cite{La-Joie:2020,Weickenmeier:2018}. We remark  that more complex models also require more informative data; either more time snapshots of tau or larger time intervals between scans.

For this first exploratory study, however, we opted for a simple model that captures the dynamics of tau propagation and allows us to establish a calibration baseline upon which more complex models can be evaluated. Recall that, although the tau propagation model is a nonlinear spatiotemporal PDE, it is parametrized by only two parameters---so, in some sense it is quite minimal. Yet, the model can reconstruct the basic features of clinical scans.
Ultimately, the reconstructed parameters $\kappa$ and $\rho$ could be used as biomarkers. Finally, the model can be run forward in time to predict tau propagation and estimate AD progression, which can serve as an overall validation. Indeed, having more than two scans would allow us to validate our model by using the first two scans to calibrate and the third scan to test the prediction.

Another challenge is the normalization of longitudinal tau PET, especially for longitudinal studies where relative change is small. The choice for a good reference region is not trivial, depends on the employed PET tracer, and is sensitive to, e.g., registration errors~\cite{Landau:2018}. The PET normalization is not a trivial matter, in fact it is an ongoing research topic~\cite{Rajagopalan:2019,Jack:2018}. Additional challenges are linked to lack of standardization in PET image acquisition protocols. Inconsistent intensity change, can be caused by several bio-physical factors such as age, weight change, and blood glucose level of the patient; or by other imaging factors such as varying scanner models or image reconstruction algorithms\cite{Boellaard:2009}. Dealing with such challenges would require cross-validation of our method with several subjects in order to select the hyper-parameters in our scheme.

\begin{landscape}
\begin{figure}
  \hspace{-0.8cm}
  \begin{tikzpicture}[
        every node/.style={anchor=south west,inner sep=0pt},
        x=1mm, y=1mm,
      ]
     \begin{scope}[shift={(0.0, -2.2cm)}]
     \node[color=anthrazit] at (-0.3cm, 0.3cm) {\small \rotatebox{90}{ \textbf{035\_S\_4114}} };
     \node (fig3) at (0,0)
       {\includegraphics[scale=0.68]{figs/adni/offset_shift_tol1E-2/\detokenize{adni_subj_035_S_4114_line_bgcorr-1.0CE}.pdf}};
     \begin{scope}[shift={(0.0, -2.0cm)}]
     \node (fig3) at (0,0)
       {\includegraphics[scale=0.68]{figs/adni/offset_shift_tol1E-2/\detokenize{adni_subj_035_S_4114_line_bgcorr-1.2CE}.pdf}};
     \begin{scope}[shift={(0.0, -2.0cm)}]
     \node (fig3) at (0,0)
       {\includegraphics[scale=0.68]{figs/adni/offset_shift_tol1E-2/\detokenize{adni_subj_035_S_4114_line_bgcorr-1.4CE}.pdf}};
     \begin{scope}[shift={(0.0, -2.0cm)}]
     \node (fig3) at (0,0)
       {\includegraphics[scale=0.68]{figs/adni/offset_shift_tol1E-2/\detokenize{adni_subj_035_S_4114_line_bgcorr-1.65CE}.pdf}};
     \begin{scope}[shift={(0.0, -2.0cm)}]
     \node (fig3) at (0,0)
       {\includegraphics[scale=0.68]{figs/adni/offset_shift_tol1E-2/\detokenize{adni_subj_035_S_4114_line_bgcorr-1.8CE}.pdf}};
     \end{scope}
     \end{scope}
     \end{scope}
     \end{scope}
     \end{scope}
   \end{tikzpicture}
   \caption{\ipoint{Sensitivity to offset for background subtraction}.
   Shown are saggital and axial cuts of tau-PET data and reconstruction for ADNI subject 035\_S\_4114. For each row, a different offset is used for the background subtraction.
   Panel a) shows axial and sagittal cuts of the raw tau-PET (for time point $T_0$); panels b) and c) show the background corrected tau scans (for time point $T_0$ and $T_1$, respectively); panels d) and e) show the initial tau seeding $c(0)$, and reconstruction $c_T$, respectively.
  }
  \label{fig:ClinPET2}
\end{figure}
\end{landscape}

\begin{landscape}
\begin{figure}
  \hspace{-0.8cm}
  \begin{tikzpicture}[
        every node/.style={anchor=south west,inner sep=0pt},
        x=1mm, y=1mm,
      ]
     \node[color=anthrazit] at (-0.3cm, 0.3cm) {\small \rotatebox{90}{ \textbf{022\_S\_6013}} };
     \node (fig1) at (0,0)
       {\includegraphics[scale=0.68]{figs/adni/offset_shift_tol1E-2/\detokenize{adni_subj_022_S_6013_line_bgcorr-1.0CE}.pdf}};
     \begin{scope}[shift={(0.0, -2.0cm)}]
     \node (fig2) at (0,0)
       {\includegraphics[scale=0.68]{figs/adni/offset_shift_tol1E-2/\detokenize{adni_subj_022_S_6013_line_bgcorr-1.2CE}.pdf}};
     \begin{scope}[shift={(0.0, -2.0cm)}]
     \node (fig2) at (0,0)
       {\includegraphics[scale=0.68]{figs/adni/offset_shift_tol1E-2/\detokenize{adni_subj_022_S_6013_line_bgcorr-1.4CE}.pdf}};
     \begin{scope}[shift={(0.0, -2.0cm)}]
     \node (fig2) at (0,0)
       {\includegraphics[scale=0.68]{figs/adni/offset_shift_tol1E-2/\detokenize{adni_subj_022_S_6013_line_bgcorr-1.65CE}.pdf}};
     \begin{scope}[shift={(0.0, -2.0cm)}]
     \node (fig2) at (0,0)
       {\includegraphics[scale=0.68]{figs/adni/offset_shift_tol1E-2/\detokenize{adni_subj_022_S_6013_line_bgcorr-1.8CE}.pdf}};

     \end{scope}
     \end{scope}
     \end{scope}
     \end{scope}
   \end{tikzpicture}
   \caption{\ipoint{Sensitivity to offset for background subtraction}.
   Shown are sagittal and axial cuts of tau-PET data and reconstruction for ADNI subject 022\_S\_6013. For each row, a different offset is used for the background subtraction.
   Panel a) shows axial and sagittal cuts of the raw tau-PET (for time point $T_0$); panels b) and c) show the background corrected tau scans (for time point $T_0$ and $T_1$, respectively); panels d) and e) show the initial tau seeding $c(0)$, and reconstruction $c_T$, respectively.
  }
  \label{fig:ClinPET3}
\end{figure}
\end{landscape}

\begin{landscape}
\begin{figure}
  \hspace{-0.8cm}
  \begin{tikzpicture}[
        every node/.style={anchor=south west,inner sep=0pt},
        x=1mm, y=1mm,
      ]
     \begin{scope}[shift={(0.0, -2.2cm)}]
     \node[color=anthrazit] at (-0.3cm, 0.3cm) {\small \rotatebox{90}{ \textbf{023\_S\_1190}} };
     \node (fig3) at (0,0)
       {\includegraphics[scale=0.68]{figs/adni/offset_shift_tol1E-2/\detokenize{adni_subj_023_S_1190_line_bgcorr-1.0CE}.pdf}};
     \begin{scope}[shift={(0.0, -2.0cm)}]
     \node (fig3) at (0,0)
       {\includegraphics[scale=0.68]{figs/adni/offset_shift_tol1E-2/\detokenize{adni_subj_023_S_1190_line_bgcorr-1.2CE}.pdf}};
     \begin{scope}[shift={(0.0, -2.0cm)}]
     \node (fig3) at (0,0)
       {\includegraphics[scale=0.68]{figs/adni/offset_shift_tol1E-2/\detokenize{adni_subj_023_S_1190_line_bgcorr-1.4CE}.pdf}};
     \begin{scope}[shift={(0.0, -2.0cm)}]
     \node (fig3) at (0,0)
       {\includegraphics[scale=0.68]{figs/adni/offset_shift_tol1E-2/\detokenize{adni_subj_023_S_1190_line_bgcorr-1.65CE}.pdf}};
     \begin{scope}[shift={(0.0, -2.0cm)}]
     \node (fig3) at (0,0)
       {\includegraphics[scale=0.68]{figs/adni/offset_shift_tol1E-2/\detokenize{adni_subj_023_S_1190_line_bgcorr-1.8CE}.pdf}};
     \end{scope}
     \end{scope}
     \end{scope}
     \end{scope}
     \end{scope}
   \end{tikzpicture}
   \caption{\ipoint{Sensitivity to offset for background subtraction}.
   Shown are sagittal and axial cuts of tau-PET data and reconstruction for ADNI subject 023\_S\_1190. For each row, a different offset is used for the background subtraction.
   Panel a) shows axial and sagittal cuts of the raw tau-PET (for time point $T_0$); panels b) and c) show the background corrected tau scans (for time point $T_0$ and $T_1$, respectively); panels d) and e) show the initial tau seeding $c(0)$, and reconstruction $c_T$, respectively.
  }
\end{figure}
\end{landscape}

\begin{landscape}
\begin{figure}
  \hspace{-0.8cm}
  \begin{tikzpicture}[
        every node/.style={anchor=south west,inner sep=0pt},
        x=1mm, y=1mm,
      ]
     \begin{scope}[shift={(0.0, -2.2cm)}]
     \node[color=anthrazit] at (-0.3cm, 0.3cm) {\small \rotatebox{90}{ \textbf{033\_S\_4179}} };
     \node (fig3) at (0,0)
       {\includegraphics[scale=0.68]{figs/adni/offset_shift_tol1E-2/\detokenize{adni_subj_033_S_4179_line_bgcorr-1.0CE}.pdf}};
     \begin{scope}[shift={(0.0, -2.0cm)}]
     \node (fig3) at (0,0)
       {\includegraphics[scale=0.68]{figs/adni/offset_shift_tol1E-2/\detokenize{adni_subj_033_S_4179_line_bgcorr-1.2CE}.pdf}};
     \begin{scope}[shift={(0.0, -2.0cm)}]
     \node (fig3) at (0,0)
       {\includegraphics[scale=0.68]{figs/adni/offset_shift_tol1E-2/\detokenize{adni_subj_033_S_4179_line_bgcorr-1.4CE}.pdf}};
     \begin{scope}[shift={(0.0, -2.0cm)}]
     \node (fig3) at (0,0)
       {\includegraphics[scale=0.68]{figs/adni/offset_shift_tol1E-2/\detokenize{adni_subj_033_S_4179_line_bgcorr-1.65CE}.pdf}};
     \begin{scope}[shift={(0.0, -2.0cm)}]
     \node (fig3) at (0,0)
       {\includegraphics[scale=0.68]{figs/adni/offset_shift_tol1E-2/\detokenize{adni_subj_033_S_4179_line_bgcorr-1.8CE}.pdf}};
     \end{scope}
     \end{scope}
     \end{scope}
     \end{scope}
     \end{scope}
   \end{tikzpicture}
   \caption{\ipoint{Sensitivity to offset for background subtraction}.
   Shown are sagittal and axial cuts of tau-PET data and reconstruction for ADNI subject 033\_S\_417. For each row, a different offset is used for the background subtraction.
   Panel a) shows axial and sagittal cuts of the raw tau-PET (for time point $T_0$); panels b) and c) show the background corrected tau scans (for time point $T_0$ and $T_1$, respectively); panels d) and e) show the initial tau seeding $c(0)$, and reconstruction $c_T$, respectively.
  }
  \label{fig:ClinPET5}
\end{figure}
\end{landscape}

\begin{landscape}
\begin{figure}
  \hspace{-0.8cm}
  \begin{tikzpicture}[
        every node/.style={anchor=south west,inner sep=0pt},
        x=1mm, y=1mm,
      ]
     \begin{scope}[shift={(0.0, -2.2cm)}]
     \node[color=anthrazit] at (-0.3cm, 0.3cm) {\small \rotatebox{90}{ \textbf{941\_S\_4036}} };
     \node (fig3) at (0,0)
       {\includegraphics[scale=0.68]{figs/adni/offset_shift_tol1E-2/\detokenize{adni_subj_941_S_4036_line_bgcorr-1.0CE}.pdf}};
     \begin{scope}[shift={(0.0, -2.0cm)}]
     \node (fig3) at (0,0)
       {\includegraphics[scale=0.68]{figs/adni/offset_shift_tol1E-2/\detokenize{adni_subj_941_S_4036_line_bgcorr-1.2CE}.pdf}};
     \begin{scope}[shift={(0.0, -2.0cm)}]
     \node (fig3) at (0,0)
       {\includegraphics[scale=0.68]{figs/adni/offset_shift_tol1E-2/\detokenize{adni_subj_941_S_4036_line_bgcorr-1.4CE}.pdf}};
     \begin{scope}[shift={(0.0, -2.0cm)}]
     \node (fig3) at (0,0)
       {\includegraphics[scale=0.68]{figs/adni/offset_shift_tol1E-2/\detokenize{adni_subj_941_S_4036_line_bgcorr-1.65CE}.pdf}};
     \begin{scope}[shift={(0.0, -2.0cm)}]
     \node (fig3) at (0,0)
       {\includegraphics[scale=0.68]{figs/adni/offset_shift_tol1E-2/\detokenize{adni_subj_941_S_4036_line_bgcorr-1.8CE}.pdf}};
     \end{scope}
     \end{scope}
     \end{scope}
     \end{scope}
     \end{scope}
   \end{tikzpicture}
   \caption{\ipoint{Sensitivity to offset for background subtraction}.
   Shown are sagittal and axial cuts of tau-PET data and reconstruction for ADNI subject 941\_S\_4036. For each row, a different offset is used for the background subtraction.
   Panel a) shows axial and sagittal cuts of the raw tau-PET (for time point $T_0$); panels b) and c) show the background corrected tau scans (for time point $T_0$ and $T_1$, respectively); panels d) and e) show the initial tau seeding $c(0)$, and reconstruction $c_T$, respectively.
  }
  \label{fig:ClinPET6}
\end{figure}
\end{landscape}



%
%

%
%
%
\bibliographystyle{spmpsci}
\interlinepenalty=10000
\bibliography{bib/literature.bib,./../tex/lib_alzh.bib}
%



\end{document}